# Ferroelectricity in 6 Å-Thick Two-dimensional $Ga_2O_3$

Tong Jiang[1,2,4,*], Han Chen[1,2,4,*], Yubo Yuan[1,2,4,*], Xiang Xu[1,2,4,*], Junwei Cao[1,2,4], Hao Wang[2,4], Xuechun Sun[1,2,4], Junshuai Li[2,4], Yaqing Ma[1,2,4], Huaze Zhu[2,4], Wenbin Li[2,4], Wei Kong[2,3,4,5]

[1]School of Materials Science and Engineering, Zhejiang University, Hangzhou 310027, China.
[2]Key Laboratory of 3D Micro/Nano Fabrication and Characterization of Zhejiang Province, School of Engineering, Westlake University, Hangzhou 310030, China.
[3]Research Center for Industries of the Future, Westlake University, Hangzhou 310030, Zhejiang, China.
[4]School of Engineering, Westlake University, Hangzhou, Zhejiang 310030, China.
[5]Westlake Institute for Optoelectronics, Hangzhou 311421, China.
*These authors contributed equally to this work.

**Abstract:**

Atomic-scale ferroelectric thin films hold great promise for high-density, low-power applications but face stability and voltage scaling challenges at extreme thinness. Here, we demonstrate ferroelectricity in single-crystalline two-dimensional (2D) $Ga_2O_3$, an ultra-wide-bandgap semiconductor, at just 6 Å thickness, exhibiting exceptional retention and thermal stability. We show that epitaxial β-$Ga_2O_3$ can be exfoliated down to a half-unit cell thickness via a self-limiting mechanism, enabling a biaxial strain-induced phase transition into a novel ferroelectric layered structure. Strain modulation enables the reduction of polarization switching voltage to 0.8 V, meeting CMOS voltage scaling requirements. Theoretical calculations reveal that switching is driven by covalent bond reconstruction, effectively countering depolarization and enhancing stability. Additionally, we integrate ferroelectric 2D $Ga_2O_3$ onto silicon using a low-temperature, back-end-of-line-compatible process. This work advances the exploration of sub-nanometer ferroelectrics, paving the way for high-density, low-power, non-volatile applications seamlessly integrated with advanced silicon technology.

**Introduction**

The exploration of stable ferroelectricity in ultrathin films holds significant promise for the development of nanoscale sensors and actuators[1-3], particularly due to the advantages of miniaturization, which facilitate low-power, high-density, non-volatile memory, and neuromorphic devices[4-6]. Despite early research on ferroelectric perovskite materials, achieving reliable ferroelectric polarization reversal at extreme thicknesses remains a persistent challenge, particularly due to size effects[7,8]. As the thickness of these films approaches a few nanometers, the depolarization field caused by uncompensated surface charges screens the ferroelectric polarization, rendering it ineffective[9-11].

In recent years, several material systems, such as $BiFeO_3$ (BFO)[12], $Hf_{0.8}Zr_{0.2}O_2$ (HZO)[13], $ZrO_2$[14], and $Bi_{1.8}Sm_{0.2}O_3$ (BSO)[15], have demonstrated the ability to maintain ferroelectric order at nanometer scales, with macroscopic ferroelectric responses observed in BSO films as thin as 1 nm. However, these materials experience varying degrees of degradation in ferroelectric retention as thickness decreases. For instance, in traditional perovskite materials like BFO, reducing the thickness down to nanometer scale results in significantly retention degradation[16]. This limitation arises because ferroelectric switching in these systems is driven by small atomic distortions mediated by long-range Coulomb interactions[7]. Furthermore, the ability to integrate ferroelectric films with silicon substrates is critical for high-density electronic devices, but the high annealing temperatures required for BSO (600 °C) and HZO (500 °C) hinder compatibility with silicon-based back-end-of-line (BEOL) processes[13,15]. Additionally, ultrathin ferroelectric films typically exhibit a large polarization switching voltage, primarily due to the increase of surface and the hindrance of domain switching by surface pinning[17]. Nevertheless, the operating voltage of ultrathin ferroelectric films reported remains incompatible with low-voltage logic circuit design requirements of complementary metal-oxide semiconductor (CMOS) technology at advanced nodes (<1 V for 14 nm nodes)[18]. Therefore, to achieve operating voltage scaling while preserving the polarization stability presents a significant challenge in the development of CMOS-compatible ultrathin ferroelectric films.

$\beta$-$Ga_2O_3$ is an emerging ultra-wide-bandgap semiconductor with exceptional electrical properties, such as low on-resistance and high breakdown field[19]. It offers commercially available large-diameter single-crystal substrates[20], similar to silicon, making it a strong candidate to rival SiC and Si in high-performance, cost-effective power electronics[21]. In its thermal dynamically stable monoclinic phase, $\beta$-$Ga_2O_3$ has a centrosymmetric structure and is traditionally not regarded as a ferroelectric material. However, it is speculated that at nearly single unit cell (UC) thickness, it could potentially transform into 2D crystal structures facilitated by electroacoustic coupling at nanoscale thinnest. These 2D-$Ga_2O_3$ films are predicted to exhibit novel electronic phases, including significantly enhanced electron mobility[22], an indirect-to-direct bandgap transition[23], and even the emergence of ferroelectricity[24]. The (100) surface of $\beta$-$Ga_2O_3$ exhibits relatively low surface energy, facilitating film cleavage along this plane[25]. However, the covalent bonding between atomic planes in $\beta$-$Ga_2O_3$ has posed a significant challenge to thinning bellow 20 nm[26]. While alternative approaches, such as oxidizing liquid Ga droplets, can produce nanometer-thick films, these are typically amorphous or polycrystalline, which prevents the emergence of macroscopic structural or electronic phases[27-29]. Therefore, the synthesis of single-crystal 2D-$Ga_2O_3$ remains a crucial but elusive goal.

In this work, we report the successful synthesis of high quality 2D-$Ga_2O_3$ single-crystalline thin films, achieving controllable thinning down to an unprecedented half-unit cell thickness, corresponding to 6 Å, through precise cleaving at the homoepitaxial interface. Our results indicate that the freestanding thin film remains stable in the monoclinic β-phase but undergoes a phase transition to the zincblende structure under biaxial strain. This transformed 2D-$Ga_2O_3$ exhibiting macroscopic ferroelectric behavior even at sub-nanometer thicknesses. Density Functional Theory (DFT) calculations confirm that stable ferroelectric polarization and switching in this system are driven by the reconfiguration of Ga-O covalent bonds and long-range distortions of oxygen atoms (~183.5 pm). This unique covalent bond reconfiguration effectively overcomes size-related limitations[30], enabling stable polarization at atomic thicknesses for extended periods. Moreover, we achieve switching voltage scaling to 0.8 V in 0.6 nm FE 2D-$Ga_2O_3$ thin films, simultaneously meeting the requirements for low operating voltage and high stability at extreme thinness. Utilizing thin film transfer techniques, we demonstrate the integration of ferroelectric 2D-$Ga_2O_3$ with silicon platforms at low temperature of 100°C, meeting BEOL requirements. These findings open new avenues for exploring extreme nanoscale ferroelectric devices and the integration of semiconductor-based ferroelectricity with advanced silicon CMOS technology for high-density integrated circuits.

**Synthesis of 2D-$Ga_2O_3$ down to half unit cell thick (6 Å)**

The monoclinic structure of β-$Ga_2O_3$ consists of four layers of Ga atoms per UC (Fig. 1a), with the (100) B surface being the easiest to cleave due to its low bonding energy. However, direct cleaving from the bulk β-$Ga_2O_3$ (100) surface results in uneven thickness, making it challenging to achieve nanometer-scale films (Extended Data Fig. 1a). In this study, we found that the homoepitaxial interface of the β-$Ga_2O_3$ (100) plane exhibits reduced binding energy due to the introduction of interfacial stress (Extended Data Fig. 1b). This led us to develop a self-limited cleaving technique—similar to methods used for 2D materials[31,32], that allows the selective separation of β-$Ga_2O_3$ epitaxial films from their substrates at the epitaxial interface (Extended Data Fig. 1c and Methods).

Using molecular beam epitaxy (MBE), we grew high-quality β-$Ga_2O_3$ homoepitaxial thin films with precise control over the epitaxial thickness. This enabled the large-area exfoliation of β-$Ga_2O_3$ films with thicknesses dictated by the epitaxial growth time. Microscopy images (Fig. 1b) display exfoliated thin films on a temporary metallic handling substrate, showing color variations that correspond to thicknesses ranging from 1 nm to 200 nm due to optical interference effects. Atomic force microscopy (AFM) measurements confirmed that the film thickness precisely matched the epitaxial thickness, revealing atomically flat surfaces with a root mean square (RMS) roughness of 0.113 nm (Extended Data Fig. 2a, b), indicative of uniform cleaving at the epitaxial interface. Cross-sectional high-angle annular dark-field scanning transmission electron microscopy (HADDF-STEM) images confirm the single-crystalline nature of freestanding films with thicknesses of 3, 1.5, and 1 UC (Fig. 1c-e).

A single UC of β-$Ga_2O_3$ contains two (100) B planes. Due to the lower surface energy of these planes, atomic incorporation preferentially occurs at step edges, leading to a half-layer-by-half-layer growth mode. Using an in-situ monitoring system with reflection high energy electron diffraction (RHEED), we controlled the epitaxial film thickness down to 0.5 UC (Extended Data Fig. 2c, d), corresponding to a thickness of 6 Å. Atomic model schematics (Fig. 1f) show that for films thicker than 0.5 UC, Ga atoms form a side-by-side six-membered ring structure projected along the [110] direction. In contrast, a film with 0.5 UC thickness projects only half as many Ga atoms in rhombic configuration. This difference is further demonstrated in plan-view transmission electron microscopy (TEM) (Fig. 1g, h), where the 1 UC film shows a hexagonal honeycomb pattern formed by Ga from upper and lower atomic planes, while the 0.5 UC film exhibits a rhombic pattern formed solely by half Ga atomic planes. These TEM images confirm the successful synthesis of 0.5 UC β-$Ga_2O_3$. Furthermore, the 0.5 UC thick freestanding β-$Ga_2O_3$ film at its 2D limit remains air-stable and retains its beta phase, countering previous theoretical predictions[33,34].

**Ferroelectricity characterization of ultrathin 2D-$Ga_2O_3$**

As β-$Ga_2O_3$ is reduced from bulk to freestanding 2D thin films, it retains its monoclinic β-phase, which exhibits inversion symmetry and, therefore, does not inherently support ferroelectricity. To our surprise, we observed novel ferroelectric (FE) behavior in such 2D-$Ga_2O_3$ films at nanometer thicknesses on a supporting substrate, persisting down to an extreme thinness of 6 Å. Figure 2a shows the amplitude and phase hysteresis loops from piezoelectric force microscopy (PFM) on a 0.5 UC film corresponding to a thickness of 6 Å, revealing characteristic signatures of ferroelectricity.

To confirm this ferroelectric behavior, we conducted multiple verification tests. First, second harmonic generation (SHG) measurements indicated a loss of inversion symmetry in the nanometer-scale 2D-$Ga_2O_3$ films on a supporting substrate (Extended Data Fig. 4), a necessary condition for ferroelectricity. Second, PFM poling tests demonstrated domain writing capabilities in the 0.5 UC film. The "box-in-box" patterns produced by applying alternating positive and negative voltages indicate reversible polarization switching (Fig. 2b). The outer unwritten region exhibited the same amplitude and phase as the central box, confirming uniform spontaneous polarization across the film. Additionally, the surface morphology remained unchanged after switching, highlighting the structural stability of the film (Extended Data Fig. 5a). The successful writing of complex patterns further demonstrated the controllability of ferroelectric domains in these ultrathin films (Extended Data Fig. 5b).

To rule out potential artifacts, we conducted additional tests. By varying the applied voltages ($V_{dc}$ and $V_{ac}$), we observed a voltage-dependent phase hysteresis loops change, thereby eliminating the possibility of artifacts due to static electricity between the PFM tip and the film surface (Extended Data Fig. 3a-c)[35]. We also measured the

PFM coercive behavior under different $V_{dc}$ frequencies to exclude the effects of charge injection or ionic migration[36,37] (Extended Data Fig. 3d, e). The most compelling evidence of ferroelectricity came from the positive-up-negative-down (PUND) method[38], which demonstrated electrical hysteresis in the 0.5 UC film with remanent polarization of 7.5 uC/cm$^2$ and coercive field of 1.4 MV/mm (Fig. 2c, Extended Data Fig. 6a). This marks the first observation of polarization-electric field response in macroscopic ferroelectric characterization on sub-nanometer-thick films, providing direct evidence of ferroelectricity and its applicability in practical nano-ferroelectric devices.

For practical applications, a wide window for operating temperature is crucial. We utilized in-situ heating SHG to characterize the Curie temperature ($T_c$) of the material (Extended Data Fig. 4). The SHG intensity remain constant until 1000 K, and after which it gradually decreases and ultimately vanishes at 1373K (Fig. 2d). When cooled to room temperature, the SHG intensity reappears (Extended Data Fig. 4c), indicating that the transition from the ferroelectric phase to the paraelectric phase driven by temperature is reversible. Based on this observation, we estimate the $T_c$ of 2D-Ga$_2$O$_3$ to be approximately 1373 K[39,40], the highest reported for thin-film ferroelectric materials to date (Extended Data Fig. 4d). This record-high $T_c$ in thin ferroelectric films means the polarization of FE 2D-Ga$_2$O$_3$ film has extremely high thermal stability. Our in-situ heated PFM experiments also confirmed that ferroelectricity remained stable over 120 °C (Extended Data Fig. 4e), with the upper limit constrained only by instrumental heating capabilities. This remarkable thermal robustness ensures reliable ferroelectric functionality over an extended operational window, even under extreme conditions. Additionally, we evaluated the retention performance of the 0.5 UC film by monitoring the decay of PFM poling patterns over time (Fig. 2e and Extended Data Fig. 5c). Using a power-law decay model, ($P(t) \propto t^{-\alpha}$, where $\alpha$ is the decay exponent), we found a minimal decay coefficient of 0.039, indicating remarkable retention enhancement compared to other ultra-thin film ferroelectric materials, such as perovskites and fluorites[15,41].

Lastly, we fabricated ferroelectric tunnel junction (FTJ) using ferroelectric 2D-Ga$_2$O$_3$, showcasing its potential for practical applications (Fig. 2f and Extended Data Fig. 6b). In these devices, a metal-ferroelectric-metal structure allows current modulation between electrodes by reversing polarity, effectively altering resistance. The tunneling electroresistance (TER) gradually saturates with increasing film thickness, reaching a ratio of 10$^4$ (Fig. 2f, Extended Data Fig. 6c, d), which allows significantly reduction of energy consumption in high-density electronics while ensuring reliability (Extended Data Fig. 6e). The extrapolated retention time of over ten years further underscores the stability of ferroelectric polarization in these films, marking 2D-Ga$_2$O$_3$ as a promising material for next-generation ultrathin ferroelectric devices. Moreover, C-AFM measurements were performed on the 2D-Ga$_2$O$_3$ film surface, revealing current hysteresis loops, which further confirm the ferroelectricity of the 2D-Ga$_2$O$_3$ (Extended Data Fig. 6f).

**Mechanisms for the phase transition and polarization switching**

We observed that the ferroelectricity in 2D-$Ga_2O_3$ thin films is highly dependent on their strain state. Specifically, while freestanding 2D-$Ga_2O_3$ thin films exhibit no ferroelectricity, ferroelectric behavior emerges when the films are subjected to strain, with the ferroelectric response becoming increasingly pronounced as stress levels rise. During the exfoliation process, 2D-$Ga_2O_3$ films are separated from their epitaxial substrates along with stressor layers, such as metallic or oxide thin films. In the absence of substrate support, these stressor layers—which inherently possess tensile residual stress due to their microstructure[42]—impose compressive strain on the 2D-$Ga_2O_3$ film.

We measured the ferroelectric response of 2D-$Ga_2O_3$ films on substrates with varying stress levels, quantified using a profilometer (Extended Data Fig. 7a, b). Strain-free 2D-$Ga_2O_3$ exhibited no ferroelectricity, in stark contrast to strained 2D-$Ga_2O_3$ films, which displayed clear ferroelectric hysteresis loops. Notably, the switching voltage ($V_c$) significantly decreased as substrate-induced stress increased (Fig. 3a). Cross-sectional TEM images provided key microscopic insights (see Methods for details). In films on straining substrates, the undistorted β-phase of $Ga_2O_3$ remains intact away from the $Ga_2O_3$-substrate interface (Fig. 3b), with geometric phase analysis (GPA) confirming minimal strain in these regions (Fig. 3c). However, GPA revealed progressively increasing compressive strain in β-$Ga_2O_3$ towards the $Ga_2O_3$-substrate interface, induced by the straining substrate. This accumulated stress leads to substantial lattice distortion near the interface. Beyond a transition zone of the highly distorted region, a distinct atomic structure forms, different from any $Ga_2O_3$ phase observed previously (Fig. 3e), accompanied by partial strain release occurring due to a phase transition (Fig. 3d).

Density functional theory (DFT) calculations confirmed the characteristics of this newly identified ferroelectric phase, revealing a 2D layered structure capable of polarization switching (Fig. 3g). The monolayer FE 2D-$Ga_2O_3$ consists of five triangular atomic layers in the sequence O-Ga-O-Ga-O, connected by covalent bonds, and belongs to the *P3m*1 space group (Fig. 3g). The asymmetric positioning of the central oxygen atom spontaneously breaks centrosymmetry, resulting in two energetically degenerate states with opposite out-of-plane electric polarization, indicating a ferroelectric atomic configuration akin to that of α-$In_2Se_3$[43,44].

Our DFT calculations, incorporating dispersion corrections, identified two possible structures for FE 2D-$Ga_2O_3$: the zincblende phase (FE-ZB) and the wurtzite phase (FE-WZ). Among these, the monolayer FE-ZB configuration consistently exhibited the lowest energy in the fully relaxed state (Extended Data Fig. 7). The out-of-plane polarization of this FE-ZB phase, calculated using the Berry phase method, was 6.84 μC/$cm^2$ (Extended Data Fig. 7c), which is consistent with the results measured by the PUND test as 7.5 μC/$cm^2$ (Fig. 2c). Additionally, the atomic model of FE-ZB phase 2D-$Ga_2O_3$ derived from DFT (Fig. 3f) matches exceptionally well with the atomic positions observed in the HAADF-STEM image of the FE 2D-$Ga_2O_3$ phase (Fig. 3e), confirming the identification of FE-ZB phase.

To explore the effect of stress on the formation of the FE 2D-$Ga_2O_3$, we computed the energy difference between the β-phase and FE-ZB phase under varying lattice parameters, producing a phase diagram (Fig. 3h). The lattice constants at position I to

V (as indicated in Fig. 3e and Extended Data Fig. 8a) depict the transition pathway from the β-phase to the FE-ZB phase. As the compressive strain increases near the substrate interface, the FE-ZB phase became energetically favorable over the β-phase, signified by a transition from the blue to red region in the phase diagram.

Thus, we identified the possible origin of ferroelectricity in 2D-Ga$_2$O$_3$ films: ultra-thin 2D-Ga$_2$O$_3$ films transition from the centrosymmetric monoclinic β-phase to the non-centrosymmetric FE-ZB phase due to in-plane biaxial compressive strain induced by the substrate. Notably, PFM measurements of ferroelectric hysteresis loops in 2D-Ga$_2$O$_3$ films of varying thicknesses revealed that ferroelectric switching signals disappear entirely when the film thickness exceeds 30 nm (Extended Data Fig. 7d). This finding suggests that thinning of single-crystal β-Ga$_2$O$_3$ films to the nanometer scale is essential for the ferroelectric phase transition.

The ability of FE-ZB phase 2D-Ga$_2$O$_3$ to overcome depolarization at extreme thinness while maintaining exceptional retention performance suggests the presence of a polarization mechanism distinct from that in traditional ferroelectric materials (see Methods for details). DFT calculations reveal the mechanism of polarization switching: under an external electric field, the long-range (~183.5 pm) displacement of O atoms is accompanied by the breaking and re-bonding of Ga-O bonds (Fig. 3g and Extended Data Fig. 8). This covalent bond reconstruction-driven polarization switching mechanism results in a higher switching barrier for FE-ZB phase 2D-Ga$_2$O$_3$ films, reaching 33.9 meV/atom (Fig. 3i), which is substantially higher than those found in other ferroelectric material systems[15] (Extended Data Table 1). Further DFT analysis revealed that biaxial compressive stress can modify this switching barrier beyond simply inducing ferroelectric phase changes. Nudged elastic band (NEB) calculations show that as the compressive stress is increased from 200 MPa to 1200 MPa, the switching barrier decreases from 31.9 meV/atom to 22.8 meV/atom (Fig. 3i). which is confirmed by experimental results that the V$_c$ of the film can be tuned by adjusting the substrate stress (Fig. 3a).

The tunable operating voltage highlights the potential of FE 2D-Ga$_2$O$_3$ films for a wide range of applications. Low-power logic circuits, such as those used in neuromorphic computing and embedded systems, require ferroelectric devices to operate at reduced voltages[45,46]. Furthermore, advanced CMOS architectures increasingly demand low operating voltages[47]. A key approach to voltage scaling is minimizing ferroelectric film thickness, as the switching voltage is directly dependent on film thickness. By optimizing substrate-induced stress and reducing the thickness of FE 2D-Ga$_2$O$_3$ films to 0.6 nm, we successfully lowered the switching voltage to 0.8 V (Fig. 3a). Notably, achieving an switching voltage below 1 V renders these ferroelectric films highly compatible with the most advanced CMOS nodes, where supply voltages have fallen below 1 V beyond the 14 nm node, necessitating extreme voltage scaling[18].

Despite the inherent challenges associated with reducing film thickness—such as increased depolarization fields, unstable polarization, and shorter retention times—the high switching barrier (Fig. 3i and Extended Data Table 1) and coercive field of FE-ZB phase 2D-Ga$_2$O$_3$ effectively mitigate these issues. These outstanding properties suppress thermal disturbances, prevents random polarization flipping, and reduces polarization decay caused by interface defects. Consequently, 0.6 nm FE 2D-$Ga_2O_3$ films simultaneously achieve both low operating voltage and high stability, positioning them as strong candidates for low-voltage logic circuits in comparison to other ultrathin ferroelectric systems (Fig. 3j, k).

**Thin film integration and ferroelectric activation**

When β-$Ga_2O_3$ is reduced to ultra-thin membranes, it exhibits mechanical properties similar to those of 2D materials. By utilizing the well-established techniques for transferring 2D materials[48], we developed a process to integrate 2D-$Ga_2O_3$ with commonly used industrial substrates, such as sapphire and silicon, as well as flexible substrates like PET (polyethylene terephthalate) and PI (polyimide) (Extended Data Fig. 9a, b). The transfer process proceeds as follows: after exfoliating 2D-$Ga_2O_3$ from the epitaxial substrate, we spin-coat its surface with PPC (polypropylene carbonate), then etch away the metallic exfoliating layer. The 2D-$Ga_2O_3$, now supported by the PPC handling layer, is affixed to the target substrate, followed by solvent cleaning to remove the PPC. We successfully transferred the 2D-$Ga_2O_3$ thin film of varying thickness onto a $SiO_2$/Si substrate (Fig. 4a and Extended Data Fig. 9c). Optical microscopy showed that the transferred 2D-$Ga_2O_3$ film had a flat surface with minimal residue (Fig. 4b). HAADF-STEM image provided a cross-sectional view of the 2D-$Ga_2O_3$ film on $SiO_2$, and energy-dispersive X-ray spectroscopy (EDS) confirmed that the film adhered conformally to the substrate, without gaps, ensuring close contact (Fig. 4c).

Initially, the as-transferred film retained its non-ferroelectric β-phase and exhibited no ferroelectric response due to the absence of strain (Fig. 4d). However, we discovered that annealing the film at temperatures above 100°C could activate its ferroelectric response (Fig. 4d). We hypothesize that upon heating, the 2D-$Ga_2O_3$ film, with a higher thermal expansion coefficient ($7.8 \times 10^{-6}$ /K along the b-axis and $6.34 \times 10^{-6}$ /K along the c-axis) than that of the Si wafer ($3 \times 10^{-6}$ /K)[49,50], experiences compressive stress, triggering a phase transition to the ferroelectric state. This transition is marked by the appearance of a ferroelectric hysteresis signal (Fig. 4d). Upon cooling to room temperature, partial stress relaxation occurs, accompanied by an increase in the $V_c$, while the ordered ferroelectric phase remains stable in the 2D-$Ga_2O_3$ film (Fig. 4e). Once the ferroelectric phase is established, the film continues to display stable ferroelectric properties even after repeated heating and cooling cycles (Extended Data Fig. 9d). With an activation temperature as low as 100°C and the full chemical compatibility of 2D-$Ga_2O_3$ with the silicon-based BEOL process, this approach offers a promising pathway for developing large-area, high-density silicon-based integrated electronics.

**Conclusion**

β-$Ga_2O_3$ is a technologically significant semiconductor, recognized for its ultra-wide bandgap and scalable single-crystal substrates similar to silicon. With well-established epitaxial growth techniques, β-$Ga_2O_3$ has been widely used in the fabrication of power transistors and photodetectors. Theoretical predictions suggest that reducing β-$Ga_2O_3$ from bulk to two-dimensional form can reveal novel structural and physical properties. In this study, we successfully fabricated sub-nanometer thick, single-crystalline 2D-$Ga_2O_3$, providing a new platform to explore these unique characteristics.

Notably, we discovered that 2D-$Ga_2O_3$ undergoes a strain-induced ferroelectric phase transition, driven by interfacial stress, resulting in a FE-ZB phase with exceptional stability at atomic-scale thickness. This 0.6 nm-thick FE 2D-$Ga_2O_3$ film satisfies the stringent requirements of advanced CMOS technologies, achieving both voltage scaling and outstanding stability. Furthermore, we demonstrated the successful integration of these FE 2D-$Ga_2O_3$ films with silicon substrates under BEOL conditions. Furthermore, we demonstrated the successful integration of these FE 2D-$Ga_2O_3$ films with silicon substrates under BEOL conditions. By incorporating photodetector, field-effect transistor, and ferroelectric functionalities of 2D-$Ga_2O_3$ within a silicon-compatible platform, 2D-$Ga_2O_3$ presents significant potential for developing multifunctional devices capable of computation, memory, and sensing in a single architecture. These findings position 2D-$Ga_2O_3$ as a promising candidate for next-generation, high-density, low-power, multifunctional integrated electronic devices.

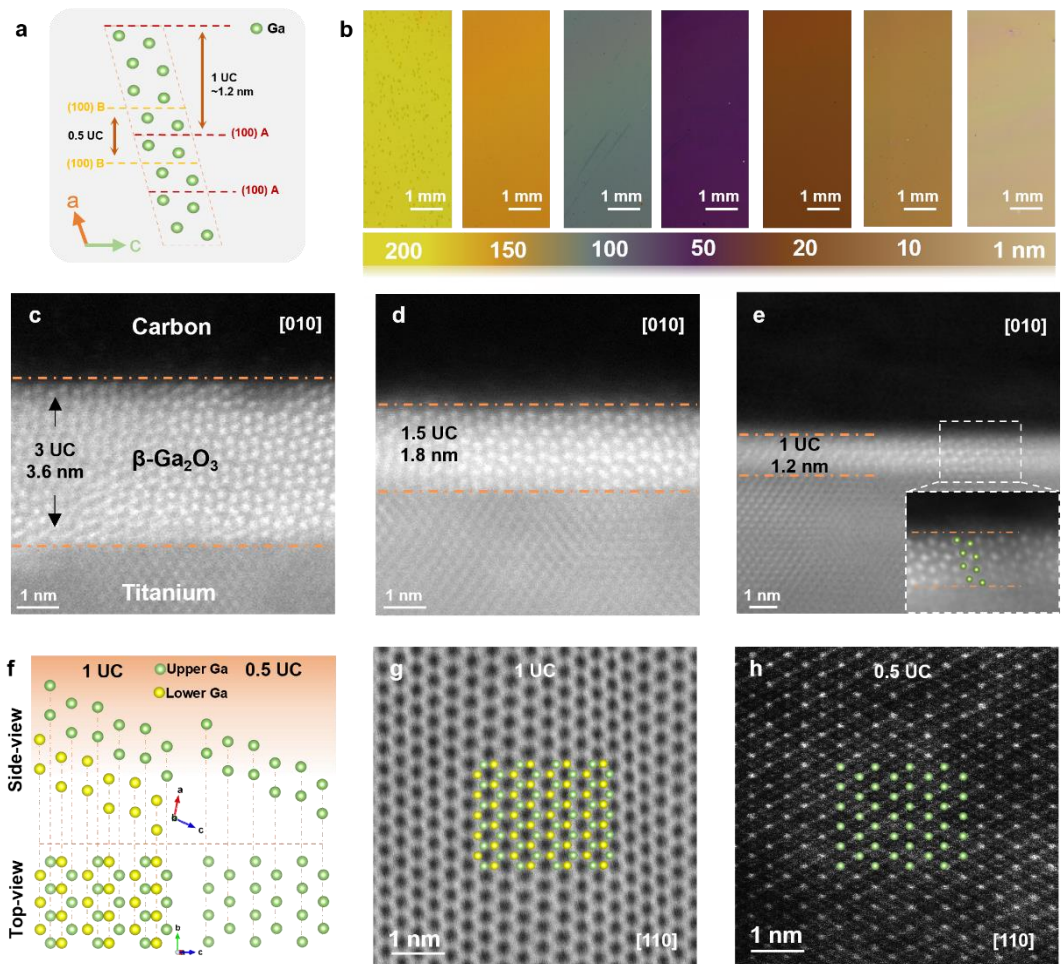

**Fig. 1 | Synthesis of single crystal β-Ga₂O₃ films with different thickness. a**, Schematic diagram of the Ga atomic structure in the β-Ga₂O₃ [010] direction. The red and yellow dashed lines represent the (100) A and (100) B planes, respectively. The thickness of one UC, composed of four Ga atomic layers, is 1.2 nm. **b**, Optical microscopy images of β-Ga₂O₃ films with various thicknesses on a metal substrate. The values on the color bar indicate film thickness. **c-e**, Cross-sectional HAADF-STEM images of β-Ga₂O₃ films with thicknesses of 3 UC, 1.5 UC, and 1 UC. The orange dashed lines denote the upper and lower interfaces of the films. **f**, Cross-sectional (side-view) and [110] projection views (top-view) of Ga atomic structures for β-Ga₂O₃ films with 1 UC and 0.5 UC thicknesses. **g** and **h**, HAADF-STEM images of the β-Ga₂O₃ (110) plane with thicknesses of 1 UC and 0.5 UC. The atomic models show the Ga configurations for each thickness as depicted in **f**.

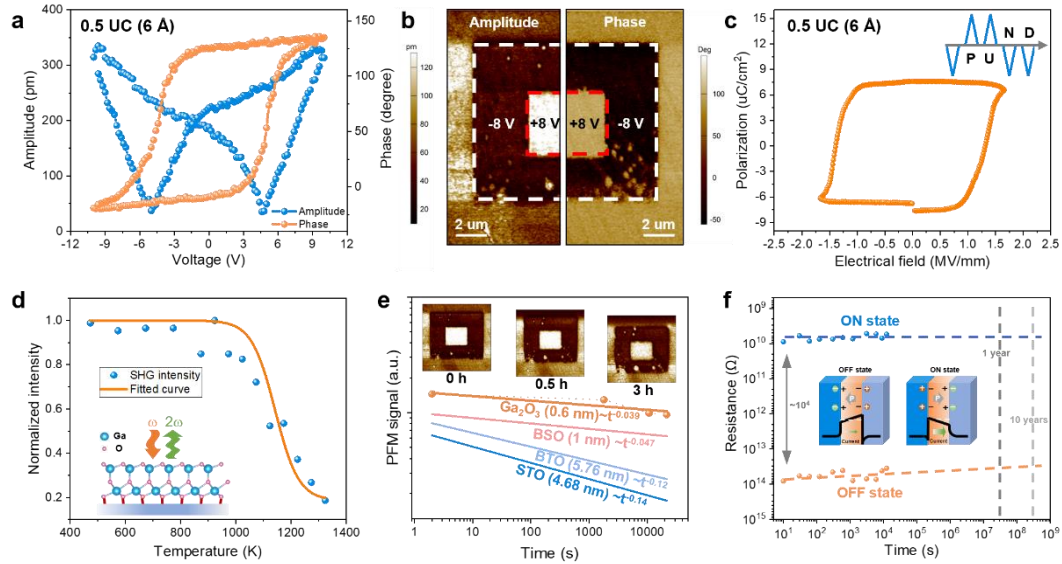

**Fig. 2 | Characterization of ferroelectricity in ultrathin 2D-Ga$_2$O$_3$ films. a**, Amplitude and phase hysteresis loops measured by PFM on the 0.5 UC thick 2D-Ga$_2$O$_3$ film. **b**, Amplitude and phase of the poling pattern written on the 0.5 UC thick 2D-Ga$_2$O$_3$ film using PFM. Negative and positive voltages of -8 V and +8 V, respectively, were applied. **c**, Polarization loop measured by PUND on a 0.5 UC thick 2D-Ga$_2$O$_3$ film. The inset shows the polarization voltage as a function of time. **d**, In-situ heating SHG testing conducted on the 3 UC-thick 2D-Ga$_2$O$_3$ film revealed the trend of SHG intensity variation with temperature. The inset shows a schematic diagram of the SHG setup. The orange and green arrows represent the normal excitation beam (ω, 1064 nm) and the generated SHG signal (2ω, 532 nm). **e**, Power-law decay fit of the PFM signal over time for the poling pattern on the 0.5 UC thick 2D-Ga$_2$O$_3$ film, with a decay exponent of 0.039. Insets display amplitude images at decay times of 0 h, 0.5 h, and 3 h. **f**, TER of the FTJ device based on 5 UC thick 2D-Ga$_2$O$_3$ film with V$_{read}$ of -0.3 V. The device demonstrates an on/off ratio exceeding 10$^4$ and excellent retention performance. The insets present schematics illustrating the polarization states and corresponding current levels in the FTJ device during the on-state and off-state.

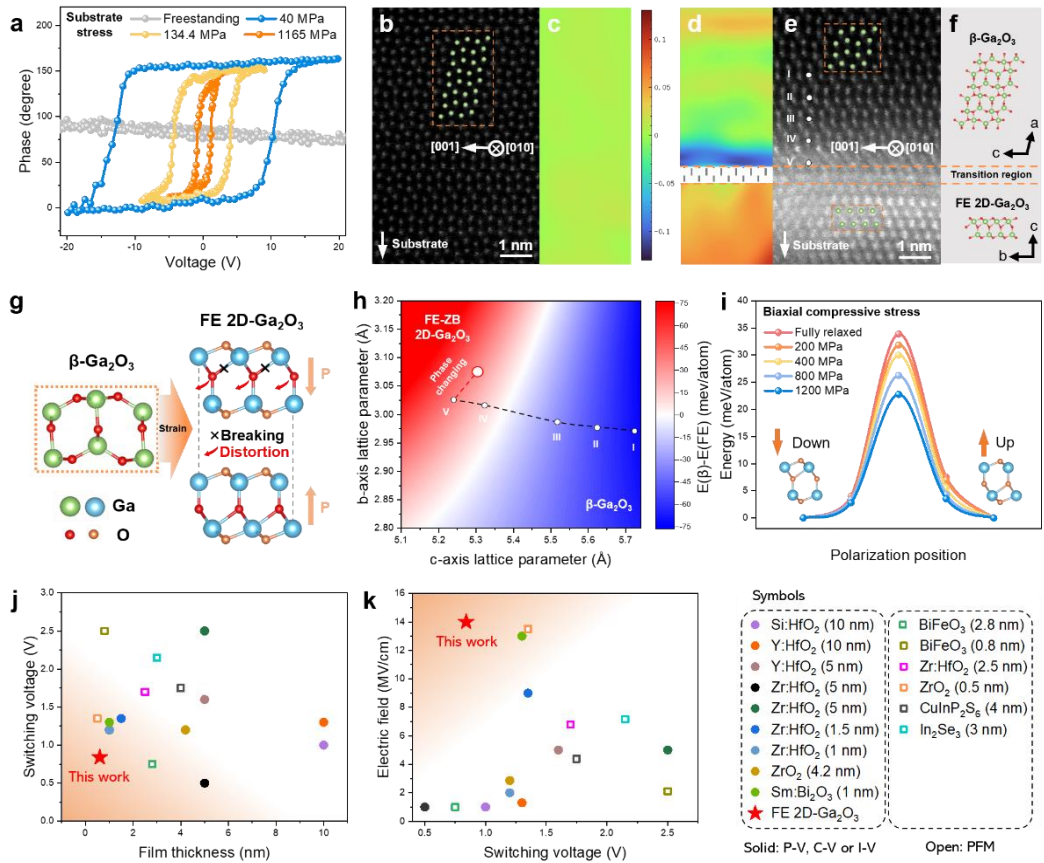

**Fig. 3 | Mechanisms of phase transition and polarization switching in FE 2D-Ga$_2$O$_3$ films. a**, PFM phase hysteresis loops measured on 0.5 UC thick FE 2D-Ga$_2$O$_3$ films under different substrate stress conditions. **b** to **e**, Cross-sectional HAADF-STEM images and GPA strain analysis of 2D-Ga$_2$O$_3$ films on a stressed substrate. Region away from the substrate interface in **b** and corresponding strain analysis in **c**, with the boxed area in **b** showing the Ga atomic positions in β-Ga$_2$O$_3$. Region near the interface in **e** and corresponding strain analysis in **d**, where the upper box in **e** highlights Ga positions in β-Ga$_2$O$_3$, and the lower box shows Ga positions in FE 2D-Ga$_2$O$_3$. The white dots labeled I–V in **e** indicate the points where the lattice parameters (c-axis) were measured and subsequently plotted in **h**. **f**, Atomic structures of β-Ga$_2$O$_3$ and FE 2D-Ga$_2$O$_3$ along the [010] direction, with the dashed area in the center indicating the transition state. **g**, Atomic models of 0.5 UC β-Ga$_2$O$_3$ and FE 2D-Ga$_2$O$_3$, with the right-side showing FE 2D-Ga$_2$O$_3$ in two polarization states. These images emphasize the distortion of central oxygen atoms during polarization switching. **h**, Energy difference contour plot between the β-phase and FE-ZB phase as a function of lattice parameter changes. The red region indicates the energetic stability of the FE-ZB phase over the β-phase, while the β-phase is more stable in the blue region. The c-axis and b-axis lattice constants for points I–V were extracted from **e** and Extended Data Fig.8a, respectively. The white dot with a red edge indicates the position of the fully relaxed 0.5 UC FE-ZB phase of 2D-Ga$_2$O$_3$. **i**, Relationship between the polarization switching barrier and biaxial compressive stress, derived from DFT calculations. Insets illustrate schematics of atomic structures in downward and upward polarization states. **j**, Benchmark of film thickness versus

switching voltage in ultrathin ferroelectric films (≤10 nm). The materials include Si:HfO$_2$[51], Y:HfO$_2$ (10 nm)[52], Y:HfO$_2$ (5 nm)[53], Zr:HfO$_2$ (5 nm)[54,55], Zr:HfO$_2$ (1.5 nm)[56], Zr:HfO$_2$ (1 nm)[13], ZrO$_2$ (4.2 nm)[57], Sm:Bi$_2$O$_3$[15], BiFeO$_3$ (2.8 nm)[58], BiFeO$_3$ (0.8 nm)[12], Zr:HfO$_2$ (2.5 nm)[59], ZrO$_2$ (0.5 nm)[14], CuInP$_2$S$_6$[60], In$_2$Se$_3$[30]. The symbols are shown on the right. The solid circles indicate that the data is derived from measurements using the standard capacitor structures (P-V, C-V or I-V), while the open squares represent data obtained from PFM measurements. Note that the PFM data may have lower accuracy due to variations in device structure and measurement parameters. **k**, Benchmark of electric field versus switching voltage in ultrathin ferroelectric films (≤10 nm). Symbols are the same as in **j**.

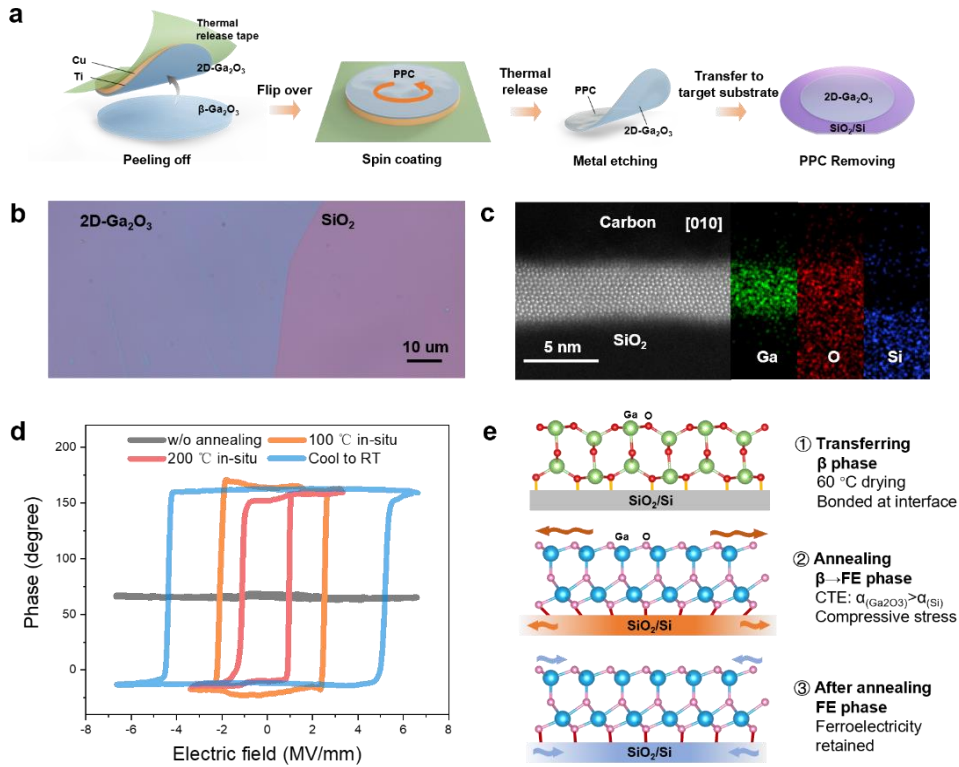

**Fig. 4 | Integration of 2D-Ga$_2$O$_3$ film with SiO$_2$/Si substrate and ferroelectric activation. a**, Schematic diagram of the 2D-Ga$_2$O$_3$ film transfer process. **b**, Optical microscope image of the 5 UC thick 2D-Ga$_2$O$_3$ film transferred onto a SiO$_2$/Si substrate. **c**, Cross-sectional HAADF-STEM image and EDS mapping of the as-transferred 2D-Ga$_2$O$_3$ film on the SiO$_2$/Si substrate. **d**, Temperature-dependent PFM phase hysteresis loops of the 5 UC thick 2D-Ga$_2$O$_3$ film on a SiO$_2$/Si substrate. **e**, Schematic diagrams illustrating the activation of ferroelectricity in 2D-Ga$_2$O$_3$ films on a SiO$_2$/Si substrate via annealing. Arrows indicate the direction of expansion or contraction in the substrate and the film.

## Methods

### Epitaxial Growth of β-Ga₂O₃ Thin Films

The β-Ga₂O₃ thin films were epitaxially grown using a plasma-assisted molecular beam epitaxy (PA-MBE) system under Ga-rich conditions at a growth temperature ($T_g$) of 650 °C. During growth, a RHEED system monitored the epitaxial quality and thickness in real time. The epitaxial substrates were unintentionally doped (UID) β-Ga₂O₃ (100) substrates, purchased from the China Electronics Technology Group Corp 46th Research Institute. Prior to epitaxial growth, the substrates underwent a standard organic cleaning process, including ultrasonic cleaning in acetone, isopropanol, and deionized water for 5 minutes each. After cleaning, the substrates were dried using dry nitrogen for 15 minutes.

### Peeling of 2D-Ga₂O₃ Thin Films

**Peeling process.** After epitaxial growth, metal stressor layers were deposited onto the surface of the β-Ga₂O₃ thin films using a magnetron sputtering system. For samples requiring a low-stress metal layer (<200 MPa), a 20 nm titanium (Ti) adhesion layer was first deposited, followed by a 1 μm copper (Cu) layer to provide the necessary stress. In contrast, for high-stress metal layer samples (>1000 MPa), an 800 nm nickel (Ni) layer was deposited to achieve the required stress levels. The β-Ga₂O₃ thin film, along with the deposited metal layer, was then simultaneously delaminated using thermal release tape (Nitto). For TEM sample preparation, indium tin oxide (ITO) was employed as the stressor layer due to its ability to induce the required stress (Extended Data Fig. 7a, b). A 300 nm ITO layer was sputtered onto the β-Ga₂O₃ epitaxial surface and subsequently peeled off using thermal release tape.

**Mechanism of Peeling.** The β-Ga₂O₃ (100) surface exhibits low surface energy, comparable to van der Waals interfaces in two-dimensional materials, which facilitates the exfoliation of thin films from bulk substrates. However, exfoliation-based fabrication of β-Ga₂O₃ thin films often results in uncontrolled and uneven thickness due to the absence of a well-defined cracking interface. Since all (100) planes within the bulk substrate are equivalent, the exfoliation process produces films with highly variable thicknesses, ranging from hundreds of nanometers to a few micrometers (Extended Data Fig. 1a). Notably, epitaxial β-Ga₂O₃ (100) experiences a reduction in bonding strength at the epitaxial interface due to the additional epitaxial stress introduced during growth. This stress can be leveraged to achieve complete delamination of the epitaxial layer precisely at the epitaxial interface, enabling the fabrication of sub-nanometer-thick films through a self-limiting exfoliation mechanism. This process is analogous to the layer-by-layer peeling observed in two-dimensional materials[31,32], allowing for the production of thin films with uniform and controllable thickness.

To implement this stress-assisted exfoliation approach, a 20 nm Ti adhesion layer was deposited, followed by a 1 μm Cu stressor layer on the epitaxial film surface. The internal stress within the Cu layer was modulated by controlling the sputter chamber

pressure during deposition[61,62]. The interface toughness relationships can be described as follows:

$$\Gamma_{Interface} < \Gamma_{GaO(100)} < \Gamma_{stressor\text{-}GaO}$$

where $\Gamma_{Interface}$ represents the toughness of the epitaxial interface, $\Gamma_{GaO\,(100)}$ denotes the toughness of the β-Ga$_2$O$_3$ (100) plane within both the bulk substrate and the epitaxial layer, and $\Gamma_{stressor\text{-}GaO}$ corresponds to the toughness between the stressor layer and the epitaxial β-Ga$_2$O$_3$ layer. In this study, ITO and Ni stressor layers—both of which induce higher stress than Cu—were also utilized as alternatives to Cu for modulating the switching barrier of ferroelectric 2D-Ga$_2$O$_3$[63].

**2D-Ga$_2$O$_3$ thin film transfer**

A mixture of polypropylene carbonate (PPC) and anisole (1:10) was spin-coated onto the surface of the 2D-Ga$_2$O$_3$ thin film at 3000 rpm for 1 minute. The film was cured at 80 °C for 3 minutes on a hot plate and subsequently released from the thermal release tape by heating to 120 °C. The Cu and Ti layers were removed using a wet etching process with FeCl$_3$ and BOE solutions, respectively. The film was then transferred onto a target substrate (such as SiO$_2$/Si or Al$_2$O$_3$) and dried at 60 °C for 1 hour. To remove the PPC layer, the film was soaked in anisole solution at 60 °C for 5 minutes, followed by sequential soaking in acetone, isopropanol, and deionized water for 10 minutes each. Finally, the film was dried using a nitrogen gun for 15 minutes.

**Stress and strain analysis**

**Measurement of substrate stress.** 4-inch silicon wafers were coated with Ti/Cu (20 nm/1 μm), Ni (800 nm) and ITO (300 nm) layers respectively using a magnetron sputtering system for stress testing. The stress of the Cu layer (DC, 5.5 W/cm$^2$, 25 nm/min) and Ni layer (DC, 5 W/cm$^2$ 10 nm/min) are modulated by changing the chamber pressure (0.1 Pa~10 Pa) during the evaporation process. Previous studies have shown that variations in processing pressure during sputtering affect the flux and energy of particles impacting the thin film in the normal direction, which in turn alters the microstructure of the deposited film[42]. The stress in the film originates from interactions between these microstructures, particularly grain boundaries. It has been demonstrated that the stress in metal films like Cu and Ni can be adjusted through experimental parameters such as processing pressure[64,65]. Similarly, the stress in ITO films can be tuned through deposition parameters[66]. In this experiment, the sputtering parameters were optimized to achieve a compressive stress of 519 MPa on the substrate (DC, 0.7 W/cm$^2$, 5 nm/min, Ar$_2$:O$_2$ = 49:1), ensuring that the strain in 2D-Ga$_2$O$_3$ is retained during focused ion beam (FIB) sample preparation. A 3D profilometer (P7, KLA-Tencor) was used to measure the wafer curvature before and after deposition, and the film stress was calculated based on these measurements.

Notably, the 2D-Ga$_2$O$_3$ film on the metallic substrate in Fig. 1c-e remains in the β-phase. We believe this occurs because the Cu metal imposes minimal stress (11.6~134.4 MPa), and the damage to the metal substrate during FIB preparation of TEM samples causes partial stress release, leading the film to revert to the β phase. The ITO substrate exerts sufficient stress (519 MPa) to maintain the induced strain even after FIB cutting, allowing the phase transition to be observed.

**TEM images stress analysis.** Microscopic strain in the β-$Ga_2O_3$ epitaxial interface was measured from high-resolution TEM images using the open-source program Strain++. The HAADF-STEM image (Fig. 3e) is divided into two sections, with the transition zone serving as the boundary. Strain is calculated separately for the upper and lower portions, while the transition zone, highlighted by the dashed box, is excluded from the strain calculation due to blurred atomic images in this region.

**Surface and structure characteristic**
**Atomic force microscopy.** An AFM (Dimension ICON, Bruker) was used to characterize the surface morphology and thickness of the films, employing non-metal-coated probes (RTESP-300, Bruker) in standard tapping mode.

**High-angle annular dark-field scanning transmission electron microscope.** HAADF-STEM images were obtained using a spherical aberration-corrected transmission electron microscope (Thermo Fisher Scientific Spectra Ultra) operating at 300 kV. The probe convergence semi-angles were set to 15 mrad and 21 mrad, while the annular detector collection angle ranged from 24 to 121 mrad and from 69 to 200 mrad, respectively. All low-order aberrations were corrected to acceptable levels, achieving a spatial resolution of approximately 0.75 Å. The cross-sectional TEM sample was prepared using FIB milling with a Thermo Fisher Scientific Helios 5 UX system. Carbon (C) and platinum (Pt) were sequentially deposited on the sample surface to protect the films during the preparation process.

**Ferroelectric characteristic**
**Positive up and negative down method.** The ferroelectric polarization charge of a 0.5 UC thick film on a metal substrate (Ni 800 nm) was tested using the PUND method. A Ti/Au metal layer, with an area of 25 × 25 $\mu m^2$, was used as the top electrode, which was deposited via shadow mask. A standard capacitor structure was then formed with the bottom Ni metal layer. Two flexible tungsten (W) probes, coated with a 10 nm beryllium copper (BeCu) alloy, were used to contact the top electrode and bottom electrode respectively, forming a current loop with a measured contact area of approximately 625 $\mu m^2$. Polarization loop tests were conducted using a semiconductor parameter analyzer (4200A-SCS, Lake Shore Cryotronics) combined with 4225-PMU and 4225-RPM modules at a frequency of 10 kHz.

**Piezoelectric force microscope.** A PFM (Cypher ES, Asylum Research) with Ti/Ir-coated probes (Asylum Research) was used to evaluate the ferroelectric properties of the film and to write complex patterns using the Litho mode. Figure 2a was measured under a $N_2$ atmosphere to ensure that the sample surface remained dry and free from impurity adsorption. In-situ heating PFM measurements were also conducted using the device's temperature control module, with a temperature range from room temperature

to 250 °C.

**Conductive Atomic Force Microscopy.** A C-AFM (Cypher ES, Asylum Research) with Ti/Ir-coated probes (Asylum Research) was used to evaluate the ferroelectric properties of the film. Dual-sweep voltage scans were performed on the film surface (±9 V for 3 UC and ±10 V for 5 UC thickness films).

**Ferroelectric Tunneling Junction.** The standard capacitor structures were fabricated for the FTJ measurements. A stress metal layer (Ti/Cu or Ni) beneath the thin film as the bottom electrode. A top electrode deposited using an E-Beam system with Ti (20 nm) / Au (50 nm) with area of 25 μm × 25 μm. FTJ device was characterized using a semiconductor parameter analyzer (4200A-SCS). The endurance performance of the Cu/Ti-2D $Ga_2O_3$ (5 UC)-Ti/Au device was applied ±6 V to reset the polarization state of the film. The device current was then read at -0.5 V, cycling through the sequence 6 V → -0.5 V → -6 V → -0.5 V for 1000 cycles.

**Second Harmonic Generation.** A Raman spectra system (Alpha300R, WITec) was used to characterize the SHG properties of the 2D-$Ga_2O_3$ film with metal substrate (Ti 20 nm/Cu 1 μm) (Extended Data Fig. 4a). The incident laser wavelength was set to 1064 nm with a power of 300 mW, and the laser was incident perpendicular to the sample surface. Polarization-dependent SHG measurements were performed in polarized light mode using the same configuration. In-situ SHG testing was performed on a 3 UC thick film (transferred to a Si substrate and annealed for 6 hours at 600 °C to provide necessary stress) using a heated sample stage (Fig. 2d), equipped with a nitrogen protective atmosphere and a water-cooling system to regulate the chamber temperature. The temperature adjustment range spanned from room temperature to 1200 °C.

**Computational Methods**
**Density functional theory of ferroelectricity.** DFT calculations were performed using the Vienna Ab initio Simulation Package (VASP)[67,68]. The projected augmented wave (PAW) method[69] was employed to account for core electrons. The exchange-correlation energy was calculated using the Perdew-Burke-Ernzerhof (PBE) functional of the generalized gradient approximation (GGA)[70]. A plane-wave kinetic energy cutoff of 700 eV was employed, as determined by prior convergence tests to ensure total energy uncertainty within 1 meV/atom[24]. The energy convergence thresholds were $10^{-6}$ eV. Residual atomic forces were minimized to below 0.01 eV/Å. For Brillouin zone sampling, a 6 × 12 × 1 mesh was applied for monolayers. Ferroelectric polarization was evaluated using the modern theory of polarization, based on the Berry phase method[71]. A vacuum layer of approximately 20 Å was included for monolayer calculations. Van der Waals interactions were accounted for using Grimme's DFT-D3 correction method[72]. The activation barriers for polarization reversal were calculated using the climbing image nudged elastic band (CI-NEB) method with fixed lattice vectors[73,74].

**Stress-dependent lattice constants**. To match the in-plane compressive stress generated during sputter deposition, we computed the energy difference between the β-phase and FE-ZB phase under varying lattice parameters, producing a phase diagram (Fig. 3h). The b-axis and c-axis lattice constants of distorted $Ga_2O_3$ (points I–V) were measured from the HAADF-STEM images (Fig. 3e and Extended Data Fig. 8a), and the potential phase transition pathways were plotted on the phase diagram. The fully relaxed lattice constants of the b-axis and c-axis for 0.5 UC FE-ZB phase 2D-$Ga_2O_3$ are 3.07 Å and 5.31 Å, respectively (the white dot with red edge in Fig. 3h).

**FE 2D-$Ga_2O_3$ overcome depolarization field**

The depolarization field significantly impacts the polarization stability of ultrathin ferroelectric materials. Investigating the mechanisms that enable FE 2D-$Ga_2O_3$ films to maintain a stable ferroelectric phase at a thickness of 6 Å is crucial. One key mechanism is covalent bond reconstruction during polarization switching, similar to that observed in α-$In_2Se_3$[30,75-77], which plays a significant role in stabilizing polarization. This process involves the breaking and re-bonding of Ga-O bonds, with the distortion of oxygen atoms reaching up to 183.5 pm, significantly exceeding the distortions observed in conventional perovskite or fluorite materials. This unique switching mechanism creates an enhanced barrier to resist the depolarization field, allowing for the establishment of a single domain with spontaneous polarization without the need for a wake-up process. As a result, the $E_c$ of the FE 2D-$Ga_2O_3$ films has reached unprecedented value of 33.9 meV/atom in ultra-thin ferroelectrics, highlighting the significant role of the covalent bond reconstruction-based switching mechanism in the FE-ZB material system.

Additionally, the single-crystalline layered structure and semiconductor properties of FE 2D-$Ga_2O_3$ further enhance its retention performance[78,79]. The two-dimensional layered configuration inherently maintains a stable asymmetric structure at ultrathin thicknesses[30]. This stability arises from the interplay between surface energy, free energy, and depolarization energy, which governs the formation of domain structures in ferroelectric films[80-82]. Reduced surface energy facilitates the formation of single-domain walls, minimizes defect density, and promotes uniform spontaneous polarization and stability across the material[83-85]. Furthermore, the semiconducting nature of 2D-$Ga_2O_3$ introduces an additional mechanism for countering depolarization effects by providing free carriers that fully compensate for surface charge imbalances. This compensation effectively mitigates the influence of depolarization fields[86]. However, these advantages would be compromised in amorphous or polycrystalline structures due to inherent structural disorder, which disrupts the formation of stable ferroelectric domains. Therefore, achieving a single-crystalline semiconductor structure is a critical factor for minimizing the impact of depolarization fields, thereby enabling robust ferroelectric behavior in ultrathin FE 2D-$Ga_2O_3$ films.

**Data availability**

The datasets generated during and/or analysed during the current study are available from the corresponding author on reasonable request.

**Acknowledgements**
We thank the Instrumentation and Service Center for Physical Sciences, Westlake Center for Micro/Nano Fabrication and Instrumentation, and Instrumentation Service Center for Molecular Sciences at Westlake University for help with characterizations. The computational resource is provided by Westlake HPC Center. This work was supported by Westlake Education Foundation. The work of Y.Y. and W.L. is supported by the National Natural Science Foundation of China (NSFC) under Project No. 62374136.

**Author contributions** Material growth was performed by T.J., H.W., J.L. and Y.M. Film exfoliation process was developed by T.J. and X.S. and performed by J.C. and H.W. Film transfer process was developed by H.Z. and T.J. and performed by J.C. PFM and SHG were performed by H.C. and T.J. Device fabrication and measurement were performed by X.X. and J.C. Electron microscopy was performed by H.Z. and T.J. DFT calculation was performed by Y.Y. under the supervision of W.L. The manuscript was co-written by T.J. and W.K. The research was supervised by W.K. All authors contributed to discussions and commented on the manuscript. T.J., H.C., Y.Y., and X.X. contributed equally to this work.

**Competing interests** The authors declare no competing interests.

**Additional information**
**Correspondence and requests for materials** should be addressed to Huaze Zhu, Wenbin Li or Wei Kong.

**Submission Date: 8th November 2024**

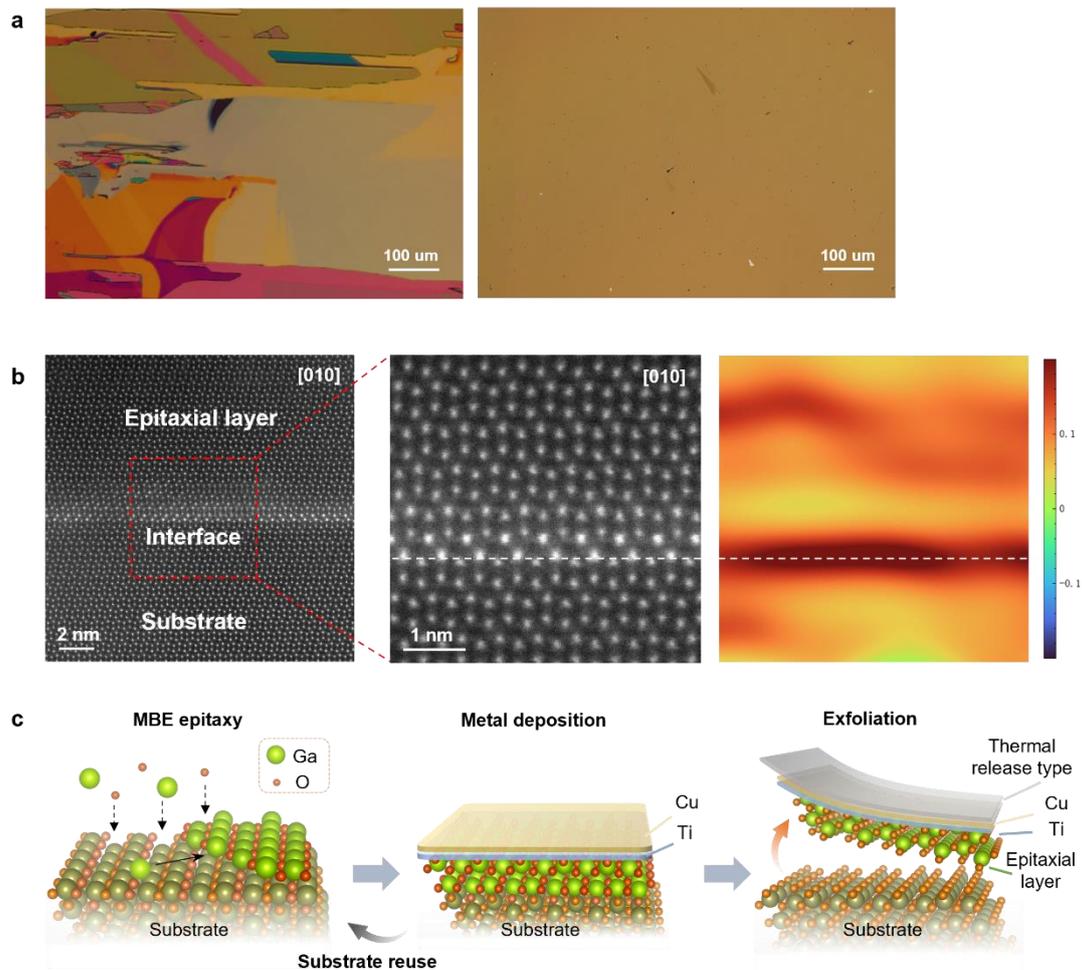

**Extended Data Fig. 1 | Mechanisms of self-limiting β-Ga$_2$O$_3$ thin film cleaving. a**, Optical microscope images of exfoliated β-Ga$_2$O$_3$ films. On the left, the film is directly exfoliated from bulk β-Ga$_2$O$_3$ with random thickness. On the right, the film is exfoliated from epitaxial β-Ga$_2$O$_3$ with uniform thickness, self-limiting at the epitaxial interface. The backside is a metal layer consisting of Ti (20 nm) and Cu (1 μm). The color contrast results from optical interference caused by the varying film thicknesses. **b**, Cross-sectional HAADF-STEM images and GPA strain analysis of β-Ga$_2$O$_3$ epitaxy on a (100) substrate. **c**, Schematic diagram of the β-Ga$_2$O$_3$ epitaxy and exfoliation process.

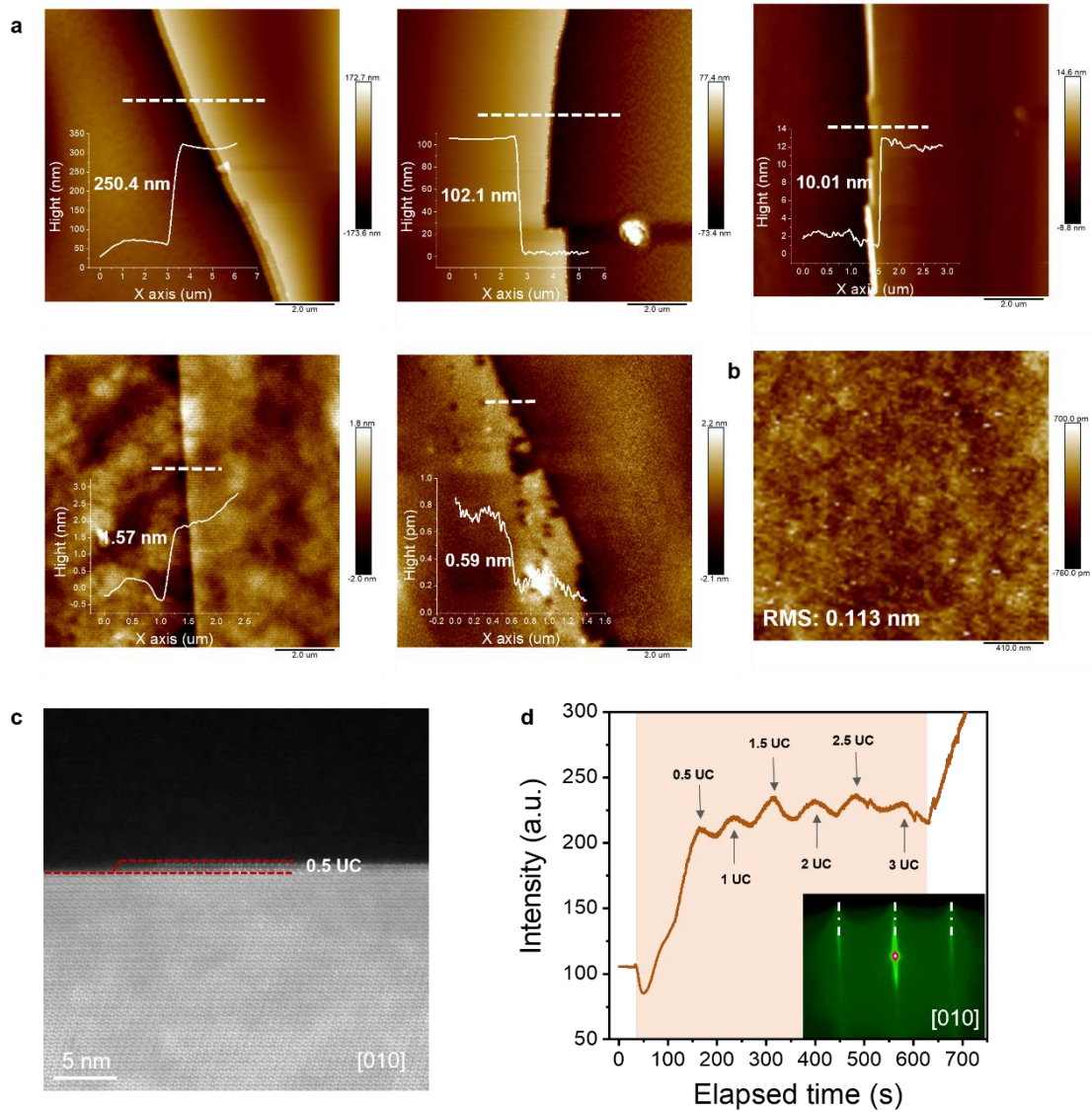

**Extended Data Fig. 2 | Thickness control of β-Ga₂O₃ epitaxial growth and exfoliation. a**, AFM characterization of β-Ga₂O₃ films with varying thicknesses. **b**, The surface morphology of the 0.5 UC thick film shows a roughness of 0.113 nm, indicating a step-free surface cleaved from a single atomic layer. **c**, Cross-sectional HAADF-STEM image of β-Ga₂O₃ epitaxy on a (100) substrate reveals a height step of 0.5 UC near the epitaxial surface, demonstrating the half-layer-by-half-layer growth of β-Ga₂O₃ on the (100) substrate. **d**, RHEED oscillation curve of β-Ga₂O₃ homoepitaxy on the (100) substrate. A single oscillation cycle corresponds to the complete growth of one layer (0.5 UC). The inset displays post-epitaxy RHEED diffraction streaks, with thin vertical lines indicating an atomically smooth epitaxial surface.

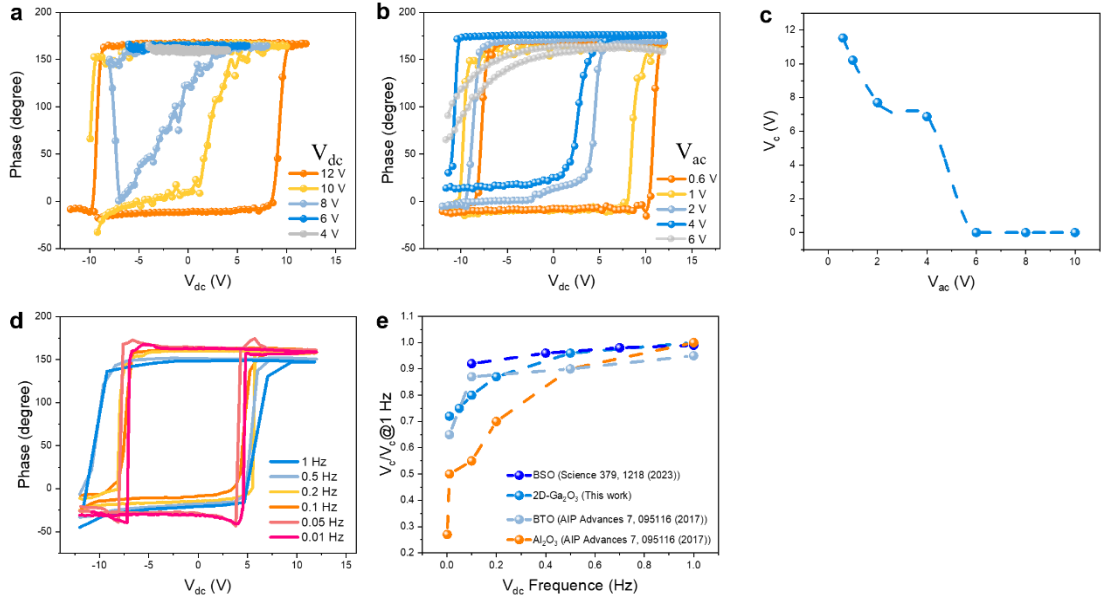

**Extended Data Fig. 3 | Verification of ferroelectricity in 3 UC thick 2D-Ga$_2$O$_3$. a**, PFM testing with varying writing voltages ($V_{dc}$) to measure the ferroelectric phase hysteresis loops of the thin films. **b**, Ferroelectric hysteresis loops measured under different reading voltages ($V_{ac}$). **c**, The relationship between $V_{ac}$ and the coercive voltage ($V_c$) of the film. As $V_{ac}$ increases, the coercive voltage decreases due to ferroelectric domain dynamics. **d**, Ferroelectric hysteresis loops measured at different $V_{dc}$ frequencies. **e**, Dependence of normalized coercive voltage ($V_c/V_c$ (1 Hz)) on the $V_{dc}$ frequency for the 2D-Ga$_2$O$_3$ film, in comparison with BSO[15], BTO[35], and Al$_2$O$_3$[35]. The weak frequency dependence of $V_c$, similar to that observed in ferroelectric materials such as BSO and BTO, excludes the possibility that the ferroelectric-like signals originate from ionic migration or charge injection, as typically observed in non-ferroelectric materials like Al$_2$O$_3$[35].

Different values of $V_{dc}$ were applied in PFM testing to measure the ferroelectric phase hysteresis loops of the thin films (Extended Data Fig. 3a). This approach was used to eliminate artifacts caused by electrostatic interactions between the probe and the film surface[35]. When $V_{dc}$ is below the coercive voltage of the film (~8 V), the polarization state of the film remains unchanged, resulting in overlapping straight lines in the hysteresis loop. When $V_{dc}$ is equal to or greater than the coercive voltage, the polarization state changes with the applied voltage, forming a clear hysteresis loop. This behavior confirms that the ferroelectric signal originates from the ferroelectric film. To further rule out potential artifacts, hysteresis loops were also measured under varying $V_{ac}$ conditions (Extended Data Fig. 3b, c). When $V_{ac}$ reaches or exceeds the coercive voltage, the polarization state of the film rapidly alternates between the "positive" and "negative" $V_{ac}$ signals, leading to the cancellation of piezoresponse[87]. If the observed signals were caused by electrostatic effects or polarization with slow switching dynamics, no clear correlation between $V_{ac}$ and the $V_c$ would be observed [88,89].

Polarization switching in ferroelectrics involves domain dynamics, including nucleation and growth processes, while in non-ferroelectrics, it typically relies on field-assisted dynamics, including charge injection, ionic movement, or electrochemical processes. Therefore, the coercive behavior observed in PFM serves as a distinguishing factor between ferroelectrics and non-ferroelectrics, as the switching time in non-ferroelectrics is strongly dependent on electric field stimulation[36]. For example, in $Al_2O_3$, at low-frequency $V_{dc}$, the switching dynamics are time-dependent, attributed to the extremely slow kinetics of charge injection or ionic redistribution[35]. In contrast, 2D-$Ga_2O_3$ films exhibit a weak dependence on $V_{dc}$ frequency (Extended Data Fig. 3d, e), similar to conventional ferroelectric materials like BTO[37] and BSO[15]. This indicates that the polarization switching in 2D-$Ga_2O_3$ films is governed by domain dynamics, thereby ruling out artifacts caused by charge injection or ionic (oxygen vacancy) migration[36].

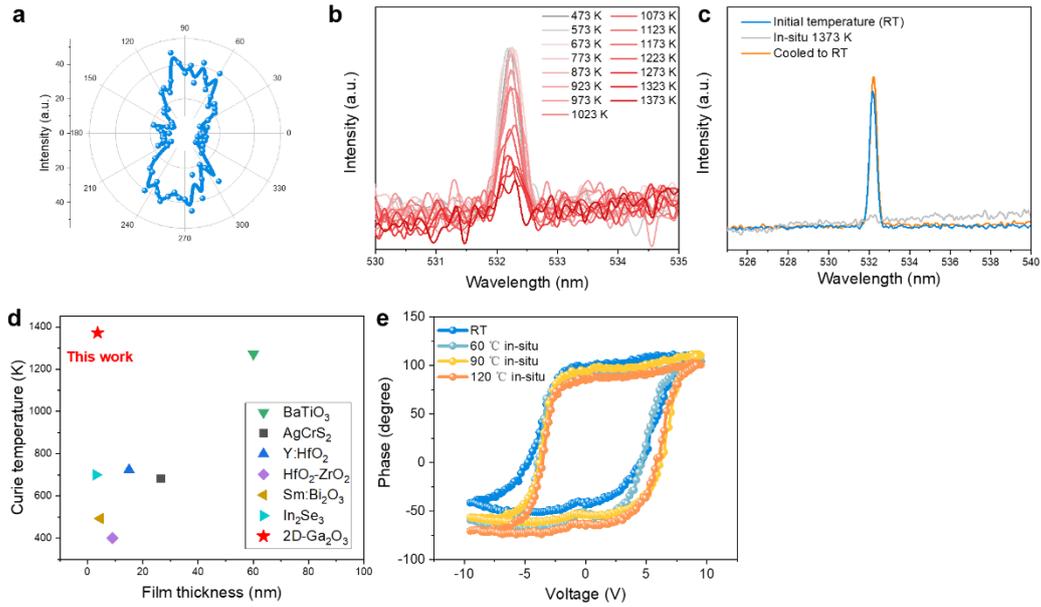

**Extended Data Fig. 4 | In-situ SHG and PFM measurements on FE 2D-Ga$_2$O$_3$. a**, Polarization-dependent SHG response under laser (1064 nm) excitation on a 0.5 UC thick 2D-Ga$_2$O$_3$ film is shown. **b**, SHG response spectra at different temperatures during the in-situ SHG testing of a 3 UC thick films. **c**, The SHG response spectra before and after in-situ heating of a 3 UC thick films. The SHG response intensity disappear at 1373 K and reappear upon cooling to room temperature, demonstrating that this process is a reversible phase transition from the ferroelectric phase to the paraelectric phase. **d**, Benchmark of film thickness versus Curie temperature in thin ferroelectric films (<100 nm). The materials include BaTiO$_3$[90], AgCrS$_2$[91], Y:HfO$_2$[92], HfO$_2$-ZrO$_2$[93], Sm:Bi$_2$O$_3$[15], In$_2$Se$_3$[30]. **e**, Phase hysteresis loops measured by PFM at different in-situ heating temperatures on 0.5 UC thick FE 2D-Ga$_2$O$_3$ film.

The generation of ferroelectric signals is fundamentally linked to the breaking of inversion symmetry in the crystal structure. SHG is highly sensitive to both the in-plane crystal symmetry and the built-in polarization of a single layer[94]. SHG exhibits ferroelectric properties because the net polarization breaks inversion symmetry, leading to a non-vanishing second-order nonlinear optical susceptibility, $\chi^{(2)}$. As a result, SHG is a powerful tool for identifying potential ferroelectric behavior in materials[95]. In particular, SHG is instrumental in characterizing the breaking of crystal symmetry, and its occurrence is typically accompanied by ferroelectric signals. Polarization-dependent SHG further reveals the symmetry of 2D-Ga$_2$O$_3$. Both ferroelectric 2D-Ga$_2$O$_3$ and α-In$_2$Se$_3$ share similar structures and space groups ($P3m1$)[43,44]. In the $P3m1$ crystal class, the monolayer contains three mirror axes, and the polarization-dependent SHG of 2D-Ga$_2$O$_3$ should exhibit a six-lobe pattern, akin to α-In$_2$Se$_3$[44]. However, studies of α-In$_2$Se$_3$ reveal a special phenomenon: the expected six-lobe SHG pattern transforms into a two-lobe pattern due to local stress from the substrate[96]. A similar reduction in symmetry due to stress has been observed in other 2D materials, such as MoS$_2$, WSe$_2$, and MoSe$_2$[97]. This occurs because layered 2D materials are highly sensitive to in-plane stress, which deforms the material and alters its structural symmetry and nonlinear photoelasticity, leading to a reduction in symmetry in the SHG patterns. Consequently, we propose that the stress from the supporting substrate induces a reduction in the structural symmetry of ferroelectric 2D-$Ga_2O_3$, resulting in the emergence of a two-lobe SHG pattern (Extended Data Fig. 4a). For verification, we determined the lattice parameters of FE 2D-$Ga_2O_3$ through HAADF-STEM images (Fig. 3e and Extended Data Fig. 8a). By measuring the atomic spacings in the [010] and [001] directions, we obtained the strained lattice constants: a = 3.025 Å, b = 5.243 Å. Based on this analysis, we confirm that the space group of 2D-$Ga_2O_3$ transitions from the relaxed *P*3*m*1 to a distorted *C*1*m*1 configuration under strain. *P*3*m*1 possesses threefold rotational symmetry along the c-axis, resulting in a six-lobe pattern in polar SHG for the relaxed FE-ZB phase. In contrast, the *C*1*m*1 lattice retains only single mirror-plane symmetry along the c-axis, leading to the emergence of a two-lobe SHG pattern."

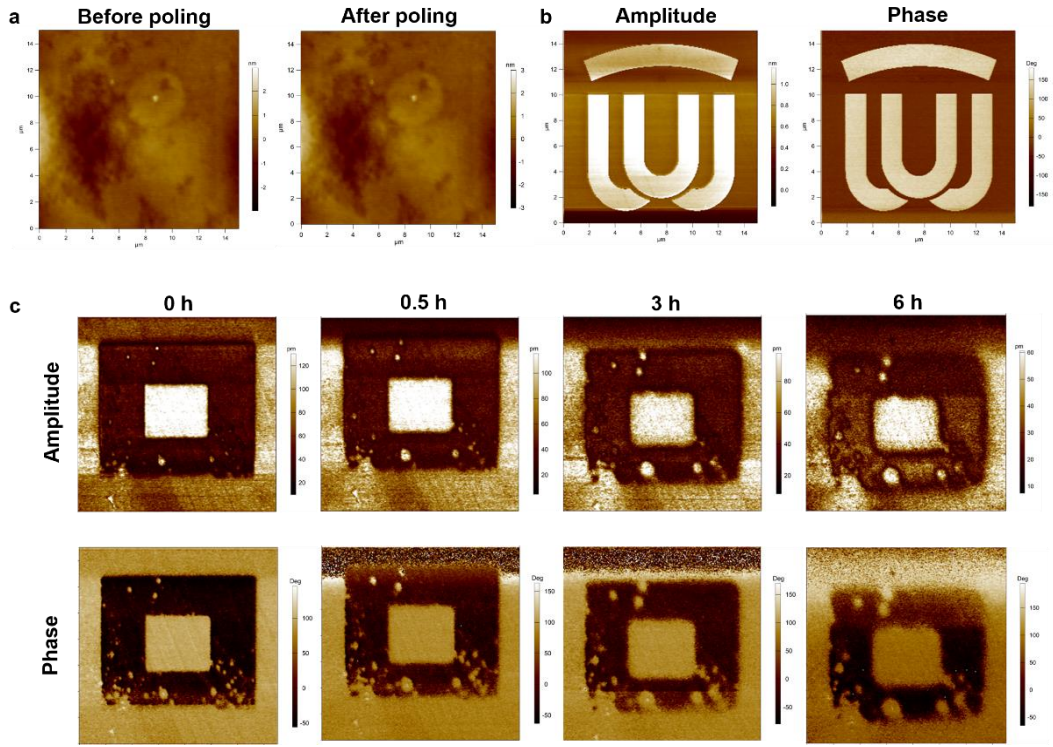

**Extended Data Fig. 5 | Characterization of PFM poling performance in FE 2D-Ga₂O₃. a**, Comparison of the surface morphology of the 0.5 UC thick FE 2D-Ga₂O₃ film before and after surface poling. **b**, Amplitude and phase images of complex patterns written onto the 0.5 UC thick FE 2D-Ga₂O₃ film, obtained using the Litho mode. **c**, Amplitude and phase images of the 0.5 UC thick FE 2D-Ga₂O₃ film at different poling decay times.

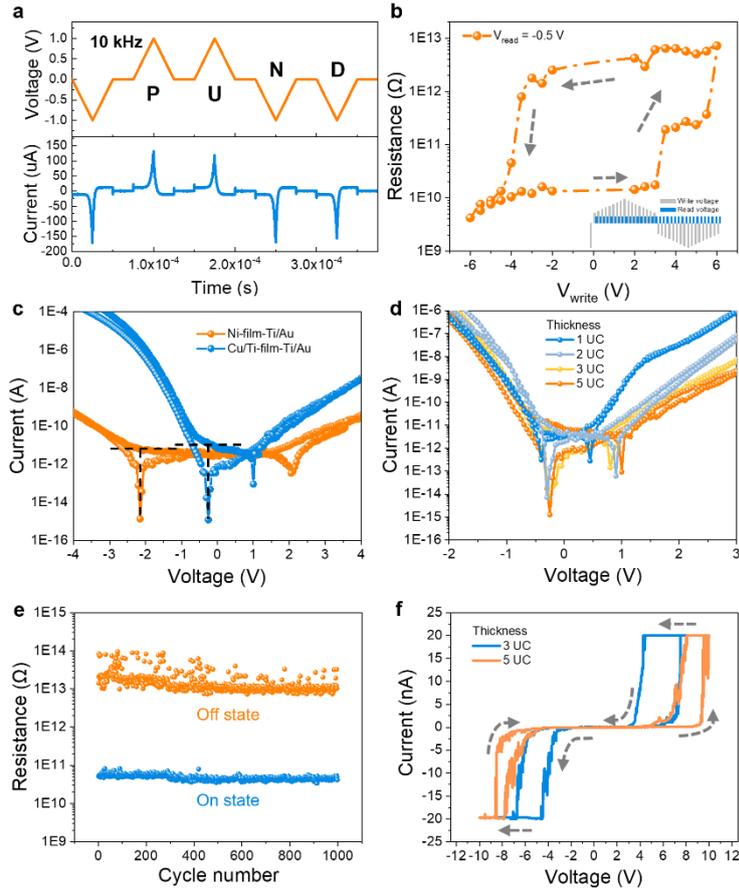

**Extended Data Fig. 6 | FE 2D-Ga$_2$O$_3$ film based PUND, FTJ and C-AFM measurements. a**, PUND pulse voltage and corresponding current waveform measured on a 0.5 UC thick ferroelectric 2D-Ga$_2$O$_3$ film at 10 kHz. **b**, Tunnel resistance of a 3 UC thick 2D-Ga$_2$O$_3$ based FTJ (V$_{read}$ = – 0.5 V) measured over a voltage pulse cycle (V$_{write}$: 0 → 6 V → – 6 V → 0 V), forming a resistance switching loop. The inset shows the applied pulse waveform. **c**, I–V switching characteristics of FTJs with a 5 UC thick 2D-Ga$_2$O$_3$ film using Ni or Cu/Ti as the top electrode; the bottom electrode in both cases is Ti/Au. **d**, I–V switching characteristics of FTJs with the structure Cu/Ti – 2D-Ga$_2$O$_3$ – Ti/Au, measured for films with thicknesses ranging from 1 to 5 UC. **e**, Endurance performance of the FTJ device (Cu/Ti – 2D-Ga$_2$O$_3$ (5 UC) – Ti/Au), evaluated by repeatedly switching the polarization state using ±6 V (Off/On state) pulses and reading the tunnel resistance at –0.5 V. **f**, Dual-sweep voltage performed using C-AFM on 2D-Ga$_2$O$_3$ films with different thicknesses (±9 V for 3 UC and ±10 V for 5 UC).

The macroscopic ferroelectric thin-film characterization method, PUND, eliminates interference from additional capacitance charging and discharging in the circuit, enabling the direct extraction of polarization charge density[38]. It provides the most direct and reliable evidence for the existence of an ordered ferroelectric phase in the film. Moreover, as the film thickness decreases, the challenge of using macroscopic PUND characterization on atomically thin ferroelectric films increases dramatically[13,14]. We successfully conducted PUND testing on FE 2D-Ga$_2$O$_3$ films as thin as sub-nanometer (6 Å), further demonstrating the strong applicability of this film in macroscopic ferroelectric devices.

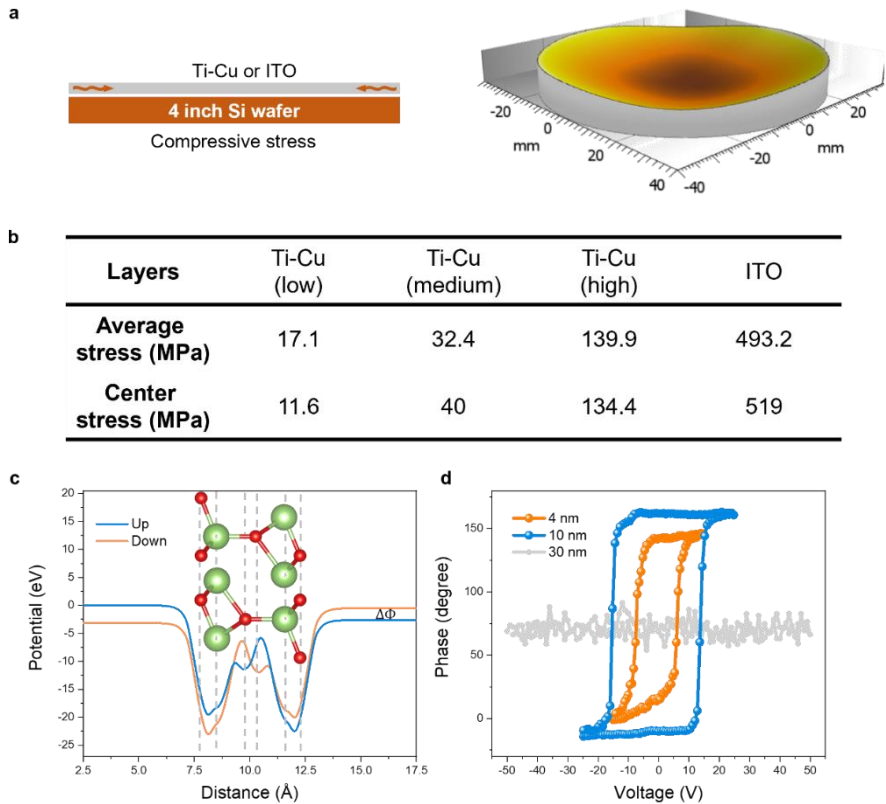

**Extended Data Fig. 7 | Strain induced Ga$_2$O$_3$ phase transition. a**, Left, structural diagram illustrating stress testing on silicon wafers. Right, illustration of strain induced by compressive stress on the silicon wafer. **b**, Measured stress data in different layers. The stress in the Cu layers was modulated by varying the chamber pressure during the evaporation process. The data in this work is subject to center stress. **c**, Polarization switching and electrostatic potentials of monolayer FE-ZB phase (fully relaxed) along the a-axis. $\Delta\Phi = 2.1$ eV, where $\Delta\Phi$ is defined as the vacuum level difference between the left and right surfaces, caused by an intrinsic dipole. **d**, PFM phase hysteresis loops measured on FE 2D-Ga$_2$O$_3$ films of varying thickness on metal substrates.

This ferroelectric phase consists of two structures: the FE-ZB and FE-WZ phases. Since FE-ZB and FE-WZ phases have similar structures and cannot be distinguished in cross-sectional HAADF-STEM, our DFT calculations show that the monolayer FE-ZB configuration has the lowest energy in the fully relaxed state (with the energy of FE-ZB being slightly lower than that of the FE-WZ phase by 0.78 meV/atom), consistent with previous calculations[24]. Additionally, our calculations indicate that the FE-WZ phase becomes unstable under biaxial compressive strain, leading to a phase transition. This phase does not correspond to the structure observed in our experiments. Therefore, we infer that the ferroelectric phase in 2D-Ga$_2$O$_3$ films predominantly arises from the FE-ZB phase.

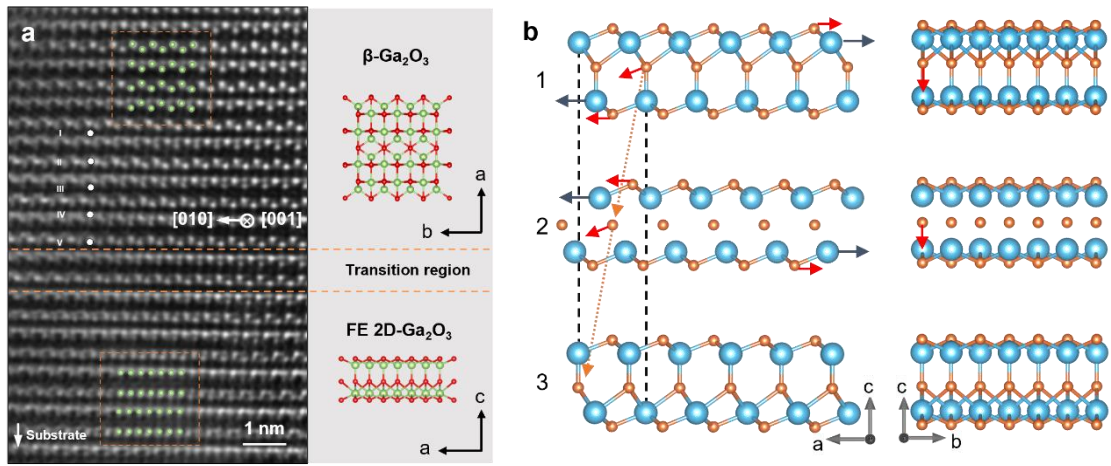

**Extended Data Fig. 8 | Cross-sectional HAADF-STEM images of 2D-Ga₂O₃ films on a stressed substrate and schematic diagrams of polarity switching in monolayer FE-ZB phase 2D-Ga₂O₃ based on DFT calculations. a**, The TEM observation direction is along the [001] direction. The upper box highlights the Ga positions in β-Ga₂O₃, while the lower box shows the Ga positions in FE 2D-Ga₂O₃, corresponding to the atomic model on the right. The white dots labeled I–V indicate the points where the lattice parameters (b-axis) were measured and subsequently plotted in Fig. 3h. **b**, Steps 1-3 illustrate the gradual atomic structural evolution during polarity switching. Blue arrows indicate the movement direction of Ga atoms, while red arrows represent the movement direction of O atoms. The dotted orange arrows indicate the positional change of the middle-layer O atoms during switching.

Polarization switching was calculated using the CI-NEB method. The primary movement occurs in the middle O atoms, which move in the a-c plane (as indicated by the dotted orange arrows in Extended Data Fig. 8b). Ga and O atoms in the upper and lower layers experience minimal displacement and return to their original positions. The distortion distance is determined based on the initial positions of the upper and lower Ga atoms. The displacement of the middle O atom is the largest, reaching approximately 183.5 pm according to calculations, accompanied by the breaking and re-bonding of Ga-O bonds.

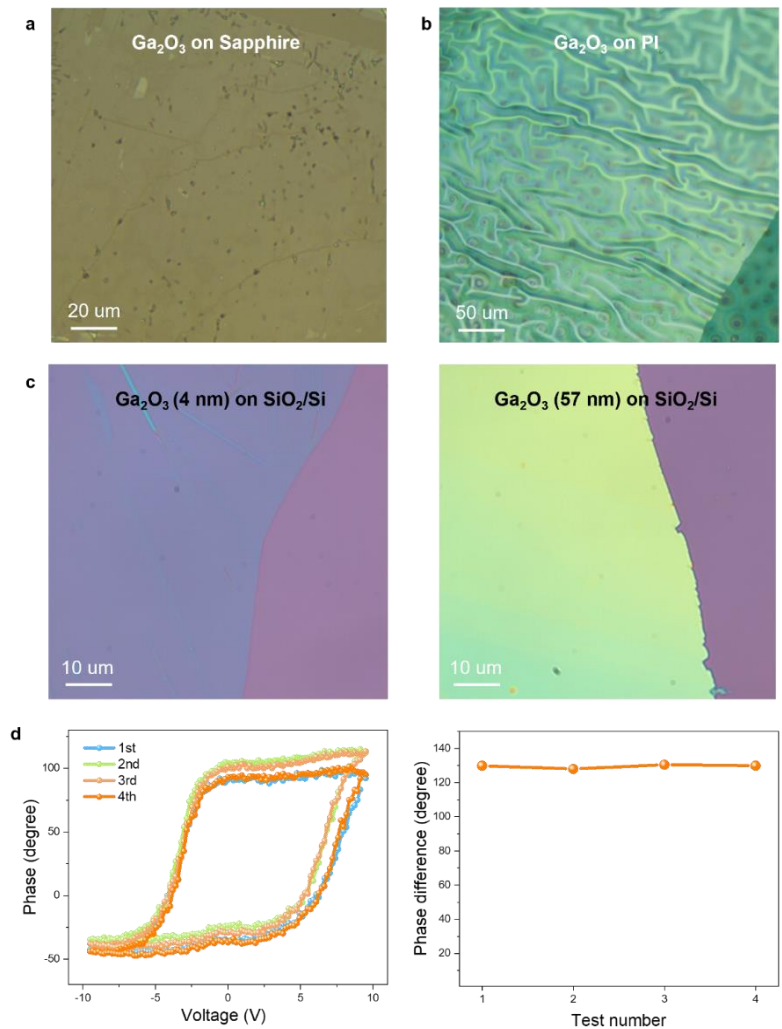

**Extended Data Fig. 9 | 2D-Ga₂O₃ transfer and thermal cycling reliability. a**. Optical microscope image of 2D-Ga₂O₃ films transferred onto sapphire. **b**, Optical microscope image of 2D-Ga₂O₃ films transferred onto flexible PI substrates. **c**, Optical microscope images of 2D-Ga₂O₃ films with varying thicknesses transferred onto SiO₂/Si substrates. **d**, PFM in-situ testing of thermal cycling in 2D-Ga₂O₃ thin films. Left: PFM phase hysteresis loops after 4 cycles from room temperature to 120 °C. The phase hysteresis loops were measured in situ at 120 °C. Right: The phase hysteresis difference between the two opposite polarizations over four cycles, demonstrating reliable polarization retention during thermal cycling.

**Extended Data Table 1 | Summary of material structures and polarization switching barriers in ultrathin ferroelectric films.**

| Materials | Space groups | Transition barriers (meV/atom) | Ref. |
|---|---|---|---|
| $BaTiO_3$ | *P*4*mm* | 1 | 98,99 |
| $\varepsilon\text{-}Fe_2O_3$ | *Pna*2$_1$ | 11.8 | 100 |
| $\alpha\text{-}In_2Se_3$ | *P*3*m1* | 13.4 | 101 |
| $Bi_6O_9$ | *Pmm*2 | 24 | 14 |
| $HfO_2$ | *Pbc*2$_1$ | 30.7~33 | 14,102 |
| 2D-$Ga_2O_3$ | *P*3*m1* | 33.9 | This work |